\documentclass[12pt]{iopart}
\usepackage{amssymb,nccbbb,cite}
\usepackage[dvips]{graphicx}
\newcommand{\sign}{\mathrm{sign}\,}
\begin{document}
\title[Algorithm for approximation of a fluxes by FSD]{Maximum likelihood algorithm for approximation of local fluctuational fluxes at the plasma periphery by fractional stable distributions}
\author{Viacheslav Saenko}
\address{Ulyanovsk State University, Leo Tolsoy str., 42, Ulyanovsk, Russian Federation, 432017}
\ead{saenkovv@gmail.com}
\begin{abstract}
Statistical properties of a local fluctuational fluxes measured at the plasma edge are investigated in the work. It's shown that the amplitudes increments of the local fluctuational fluxes decrease by power law. For approximation of experimental  PDFs the fractional stable distributions are used. The new algorithm of statistical estimation of the FSD parameters based on maximum likelihood method is described. By using of the algorithm the  parameters of FSD are estimated by using experimental samples. It is shown good agreement between experimental and theoretical distributions. On the basis of this results the conclusion is made about applicability of the CTRW model for description of a processes underlying of the plasma turbulence.
\end{abstract}


\section{Introduction}
Formation of turbulence in plasma is one of important cause  which don't let to realize a fusion. As a investigations show, the plasma turbulence possesses have a number of properties which become apparent regardless of experiment conditions and devices. First of all it was established that probability density function (PDF) of amplitudes of  plasma density fluctuation and local turbulent fluxes don't described by Gauss distribution \cite{Jha1992, Trasarti-Battistoni2002, Skvortsova2005, Carreras1999b, Carreras1996}. The empirical PDFs have heavy tails with power asymptotic $x^{-\alpha-1}$ and sharp peaks. Another property of empirical PDF is self-similarity of ones
$$
\rho(x,t)=t^{-H}\rho(xt^{-H},1).
$$
This fact was established in
works \cite{Trasarti-Battistoni2002, Carreras1999b, Hidalgo2002a, Antar2003} where by scaling transformation of PDFs was shown their identity. In works \cite{Carreras1998b, Carreras1998a} the Hurst
parameters were calculated for the particle fluctuation fluxes which were measured in three stellarators TJ-IU,
W7-AS, ATF and two tokamaks – TJ-I, JET. In the works \cite{Saenko2009, Saenko2010} have been shown that the local fluctuating fluxes possess self-similarity property and the increments distribution can be described by fractional stable distributions. In all cases, the Hurst exponent was found to be $H>0.5$ and lie in a fairly narrow range $0.62\leqslant H \leqslant 0.72$ in the plasma core.
Here $H$ is the Hurst parameter and $\rho(x,1)$ is distribution at the initial moment.

The presence of the heavy-tail distributions have resulted necessity refusal of using classical approaches with central limit theorem and Gauss distribution in the basis and development new models which be able to describe the observed dependence. In the works \cite{Batanov2002, Skvortsova2005} was proposed the model with assumption that in plasma a number of gaussian processes appear each of which has it own values of parameters. As a result the PDF of such set of gaussian processes is described by shift-scaling mixture of gaussian distributions. In the works the parameters of shift-scaling mixtures of distributions are statistically estimated with respect to experimental dates and theoretical and empirical distributions was compared each other. Such comparison gave a good coincidence theoretical and empirical distributions.

Another method is based on proposition of cascade process dissipation of kinetic energy in turbulent flow.
It can be explained by presence of various structures on different spatial scaling. This assumption leads to a
multifractal conception \cite{Antar2001, Budaev2004}. In the framework of this conception spectrum of scales can be obtained by use
of structure function method \cite{Budaev2006,Budaev2008}. In this work it was showed that superdiffusion processes are developed in
plasma.

In the present work the approach with Continuous Time Random Walk (CTRW) model \cite{Metzler2000} in the basis is described. This model was firstly described in the work \cite{Montroll1965} and further was used for description of processes in physics, chemistry, biology and economics. It also has found a use in the plasma physics for description of turbulent processes. In works \cite{SanchezBurillo2009a, SanchezBurillo2010} this model lies in the basis of study of  radial transport models of charged particles in toroidal devices. In the work \cite{Gustafson2012} shown that transport of superthermal ions in closed magnetic trap is described by Levy walks.

The CTRW model is based on the assumption that a randomly walking particle can either move or be at rest. If the PDFs of the particle jump length $\rho(x)$ and the time during which the particle is at rest $\eta(t)$ are power-law functions
\begin{eqnarray}
\rho(x)&\propto \alpha x_0^\alpha x^{-\alpha-1},\ 0<\alpha\leqslant2,\ x\to\infty,\nonumber\\
\eta(t)&\propto \beta t_0^\beta t^{-\beta-1},\ 0<\beta\leqslant1,\ t\to\infty.\nonumber
\end{eqnarray}

In the framework of this assumptions the asymptotic (at $t\to\infty$) of particle coordinate is described by fractional diffusion equation \cite{Metzler2000}
is expressed through partial derivatives of fractional orders. Equations of such kind find a use for description of propagation of ions and electrons in the plasma of closed magnetic traps  \cite{Del-Castillo-Negrete2004, Mier2008a}. The order of derivative with respect to time is defined by exponent $\beta$ and
order of derivative with respect to space coordinate is defined by exponent $\alpha$. In the particular, for $\beta=1, \alpha=2$ fractional diffusion quation is reduced to the conventional diffusion equation
$$
\frac{\partial p(x,t)}{\partial t}=D\frac{\partial^2 p(x,t)}{\partial t^2},
$$
where $D$ is the diffusion coefficient. For the initial condition $p(x,0)=\delta(x)$ the solution of this equation has the form
\begin{equation}\label{eq:gausspdf}
p(x,t)=(Dt)^{-1/2}g_G\left(x(Dt)^{-1/2};2,1\right),
\end{equation}
where $g_G(x;2,1)=(2\pi)^{-1/2}\exp(-x^2/2)$ is the Gauss distribution. It follows from the (\ref{eq:gausspdf}) that the solution is self-similar, with the Hurst exponent $H=1/2$. If  $\beta\neq1,\alpha\neq2$ the solution of fractional diffusion equation is expressed in terms FSDs \cite{Uchaikin2000,Bening2006}.

In the works \cite{Saenko2009, Saenko2010} the FSDs were applied for approximation of empirical PDFs of local fluctuational fluxes at the edge of the plasma cord. The FSD parameters have been statistically estimated according to experimental time series by using methods are described there. Comparison of the theoretical and the empirical distributions shown good agreement between them. However, to estimate the distribution parameters in the processing of time-series had to significantly reduce the amount of the original sample. The reason of this is described in \cite{Saenko2009}. As a result the sample volume is about 300-500 values, and the algorithm which was used there suppose presence a sample of an enough big size. As result the estimated values had enough big statistical error. In this work the new algorithm for statistical estimating FSD parameter is described. This algorithm base on maximum likelihood method and It is less demanding for the sample. By using this method the parameters of the FSD will estimated according to the amplitudes increments of the local fluctuational fluxes. After that the empirical and theoretical distributions will be compared together.

The work has following structure. In the section~\ref{sec:FSD} necessary theoretical  information about FSD are presented. In the section~\ref{sec:estFSD}
the maximum likelihood algorithm for statistical estimation of FSD parameter is described. In the section~\ref{sec:approxAlg} the method of processing of experimental time series is described and the section~\ref{sec:results} the results of approximation and conclusions are presented.

\section{Fractional Stable Distributions}\label{sec:FSD}

Fractional stable densities (FSDs) were first introduced in \cite{Kotulski1995} as limit distributions in the CTRW model and, independently, were used by Kolokol'tsov et al \cite{Kolokoltsov2001}. Let us consider the following summation scheme. Let $X_1, X_2,\dots$ be independent, identically distributed random variables and $T_1, T_2,\dots$ be variables that are independent of both each other and also of the sequence $X_1, X_2,\dots$, which are identically distributed over the positive semi-axis. We consider the compound process
\begin{equation}\label{eq:summ}
S(t)=\sum_{j=1}^{N(t)}X_j,
\end{equation}
where $N(t)$ is the counting process
$N(t)=\sum_{j=1}^{N(t)}T_j\leqslant t<\sum_{j=1}^{N(t)+1}T_j,\quad T_j>0.$
We assume that random variables $X_j$ belong to the region of normal attraction of a strictly stable law with characteristic function $\hat g(k;\alpha,\theta)$  ($0<\alpha\leqslant2$) and $T_j$ belong to the region of normal attraction of a one-sided strictly stable law with characteristic function $\hat g(k;\beta,1)$ ($0<\beta\leqslant 1$). The characteristic function of the strictly stable law has the form
\begin{equation}\label{eq:form_c}
\hat g(k;\mu,\theta)=\exp\left\{-|k|^\mu\exp\left(-i\frac{\pi\theta\mu}{2}\sign k\right)\right \},
\end{equation}
where $0<\mu\leqslant 2$, $|\theta|\leqslant \min(1,2/\mu-1)$ \cite{Zolotarev1986}.

The sum (\ref{eq:summ}) is physically interpreted as a random coordinate of a particle at time $t$ that undergoes random walk in the CTRW model. $X_j$ is then interpreted as a jump of the particle between two spatial points, whereas $T_j$ is interpreted as a random waiting time of the particle between two successive jumps.

In \cite{Kolokoltsov2001}, it was shown that, on the above assumptions, asymptotic (at $t\to\infty$) distribution of the sum (\ref{eq:summ})) is described by the FSD
\begin{equation}\label{eq:fsd}
q(x;\alpha,\beta,\theta)=\int\limits_0^{\infty}g(xy^{\beta/\alpha};\alpha,\theta)g(y;\beta,1)y^{\beta/\alpha}dy,
\end{equation}
where $g(x;\alpha,\theta)$ is the strictly stable probability density, whereas $g(y;\beta,1)$ is the one-sided strictly stable probability density with characteristic function (\ref{eq:form_c}).

	As already noted, the physical interpretation of the sum (\ref{eq:summ}) is the particle coordinate in the CTRW model. It can be shown (see \cite{Metzler2000}) that the asymptotic coordinate distribution is described by the equation
\begin{equation}\label{eq:commdfs}
\frac{\partial^\beta p(x,t)}{\partial t^\beta}=-D(-\Delta)^{\alpha/2}p(x,t)+\frac{t^{-\beta}}{\Gamma(1-\beta)}\delta(x),
\end{equation}
where $\partial^{\beta}/\partial t^{\beta}$ is the Riemann--Liouville fractional derivative and $(-\Delta)^{\alpha/2}$ is the fractional-order Laplacian \cite{Samko1973}. In \cite{Uchaikin2000}, it is shown that the solution to this equation has the form
\begin{equation}\label{eq:sol_commdfseq}
p(x,t)=\left(Dt^{\beta}\right)^{-1/\alpha}q\left(|x|\left(Dt^\beta\right)^{-1/\alpha};\alpha,\beta,0\right),
\end{equation}
where $q(x;\alpha,\beta,0)$ is the FSD.
	The solution (\ref{eq:sol_commdfseq}) can be represented in the form
$$
p(x,t)=(D't)^{-H}q\left( |x|\left(D't\right)^{-H};\alpha,\beta,0\right),
$$
where $D'=D^{1/\beta}$ and
\begin{equation}\label{eq:hurst}
H=\beta/\alpha.
\end{equation}
 It follows herefrom that, first, the density $p(x,t)$ possesses self-similarity and, second, the solution (\ref{eq:sol_commdfseq}) at $x\to\infty$ decreases as $x^{-\alpha-1}$.

\section{Estimator for Parameters of FSD}\label{sec:estFSD}

Let $Y_1,\dots, Y_N$ is sample of independent identical distributed random variables. Let distribution of each of them coincide with FSD \eref{eq:fsd} with characteristic parameters $\alpha$ and $\beta$. The task consist in estimation of a values $\hat\alpha,\hat\beta$ of the parameters $\alpha$ and $\beta$ by sample $Y_1,\dots,Y_N$.

It is well known that maximum likelihood estimation are parameters values $\hat\alpha,\hat\beta$ under which
the logarithmic likelihood function $L(y;\alpha,\beta)$ is reaching maximum:
\begin{equation}\label{eq:maxLHest}
\max_{(\hat\alpha,\hat\beta)}L(Y;\hat\alpha,\hat\beta)=
\max_{(\hat\alpha,\hat\beta)}\frac{1}{N}\sum_{i=1}^{N}l(Y_i;\hat\alpha,\hat\beta),
\end{equation}
where
\begin{equation}\label{eq:loglhf}
l(Y_i;\hat\alpha,\hat\beta)=\log q(Y_i;\hat\alpha,\hat\beta,0)
\end{equation}
and $q(Y_i;\hat\alpha,\hat\beta)$ is the FSD at the point $Y_i$.
As we can see in order to evaluate the maximum likelihood estimation it is necessary to be able calculate
the density at the points $Y_1,\dots,Y_N$ for given parameters values. However, in common case the
FSD densities aren't expressed through elementary function. Therefore Monte-Carlo method is used for calculation of PDF
of FSD.

The task of calculation
density of symmetric ($\theta=0$) FSD was considered in the work \cite{Uchaikin2002} and the following estimator was obtained
\begin{equation}\label{eq:estfsd}
\hat q(x;\alpha,\beta,0)=\frac{1}{M}\sum_{j=1}^M\frac{[S_j(\beta)]^{\beta/\alpha}}
{\sqrt{4\pi S_j(\alpha/2)}}
\exp\left\{\frac{x^2[S_j(\beta)]^{2\beta/\alpha}}
{4S_j(\alpha/2)}\right\},
\end{equation}
where $S_j(\alpha/2)$ and $S_j(\beta)$ are independent identical distributed random variables either have one-sided
strictly stable distribution with characteristic parameters $\alpha/2$ and $\beta$ respectively.
The Kanter's algorithm \cite{Kanter1975} was used for simulation of ones.

The estimator \eref{eq:estfsd} allow us to estimate PDF at given point. As a result an obtained value  don't have systematic component of an error. Moreover if we have set of points $x_1,x_2,\dots, x_n$, we can simultaneously estimate PDF at all given points for one realization of random variables $S_j(\alpha/2)$ and $S_j(\beta)$. It is noticeable decrease calculation volume and magnitude of statistical error in comparison with histogram estimation of PDF.

Let us use the estimator \eref{eq:estfsd} for evaluation of likelihood function. Substitute the estimator in \eref{eq:loglhf}
we obtain the Monte Carlo estimator for likelihood function
\begin{equation}\label{eq:lhf}
\hat l(Y_i;\alpha,\beta)=-\ln M
+\sum_{j=1}^M\ln\left[\frac{[S_j(\beta)]^{\beta/\alpha}}
{\sqrt{4\pi S_j(\alpha/2)}}\exp\left\{\frac{Y_i^2[S_j(\beta)]^{2\beta/\alpha}}
{4S_j(\alpha/2)}\right\}\right].
\end{equation}
Now the task consist in maximization of the likelihood function by parameters $\alpha,\beta$.

It is necessary to use direct search methods for functions of $n$-variables in the task of
likelihood function maximization. In the present work the Hooke-Jeeves's method \cite{Bunday1984} was applied.
Presence of any additional information about position of a maximum of function allow considerable to decrease a
search time. This information will allow to set initial position of basic point from which the maximization algorithm will be starting.
The nearer this point is placed toward the true value, the  smaller time will necessary for maximization of function. It is
possible to use an algorithm of statistical estimation of FSD parameters which was described in
\cite{Bening2004} for determination initial basic point.  The estimators for parameters of FSD
are presented in the Appendix~\ref{sec:prmestmom}. As a result a position of initial point may be determined by formulas
\eref{eq:theta_est} -- \eref{eq:lambda_est}.

It is necessary to take into account the fact that domain within which a maximum of a function is being searched is bounded.
This domain is defined by domain of variation of parameters of FSD
\begin{equation}\label{eq:prmdomain}
G=\{(\alpha,\beta,\lambda): 0<\alpha\leqslant2,0<\beta\leqslant1,\lambda>0\}.
\end{equation}
Beyond the bounds of this domain a density of FSD isn't defined.

It is convenient to use of helper function for searching of a maximum of likelihood function under conditions \eref{eq:prmdomain}
\begin{equation}\label{eq:lhfpen}
\mathcal{L}(x;\mathrm{P})=\left\{\begin{array}{cc}
l(x;\mathrm{\tilde P})-e^{|\alpha-\tilde\alpha|+|\beta-\tilde\beta|}+1,& \mathrm{P}\notin G,\\
l(x;\mathrm{P}), &\mathrm{P}\in G.
\end{array}\right.
\end{equation}
Here the points $\mathrm{P}=(\alpha,\beta)$, and $\mathrm{\tilde P}=(\tilde\alpha,\tilde\beta)$, where
$\tilde\alpha,\tilde\beta,\tilde\lambda$ are bounding values domain $G$ for corresponding parameter. The
function $\mathcal{L}(x;\mathrm{P})$ redefine likelihood function $l(x;\mathrm{P})$ to domain
$G'=\{(\alpha',\beta',\lambda'):-\infty<\alpha'<\infty,-\infty<\beta'<\infty,-\infty<\lambda'<\infty\}$
and it is penalty for exit of algorithm out of the domain $G$. This function is decreasing the value of the function $l(x;\mathrm{P})$
if point $\mathrm{P}\notin G$. At the same time such decreasing the stronger, the farther the point $\mathrm{P}$
exceed the bounds of the domain $G$. If the point $\mathrm{P}\in G$ then $l(x;\mathrm{P})$ is not changing. Thus,
the function \eref{eq:lhfpen} decrease a value of likelihood function if the algorithm exceed the bounds of the $G$ domain.
This not allow to algorithm to move further in this direction. As a result this lead to returning the algorithm back
into the $G$ domain.

\section{Approximation of turbulent particle fluxes}\label{sec:approxAlg}

Now we apply the FSD to describing probability densities of amplitudes of turbulent particle fluxes.

Turbulent particle fluxes were measured in the edge plasma of the L-2M stellarator \cite{Meshcheryakov2005} with the use of a set of three Langmuir probes. It is well known that the flux $\tilde\Gamma_j$ is defined as,
$$
\tilde\Gamma_j=\delta n_j\delta v^r_j, \qquad j=1\dots N,
$$
where $\delta n_j$ are plasma density fluctuations, $\delta v^r_j$ are fluctuations of the radial velocity, and $N$ is the number of points in the time sample being processed. The fluctuations of the radial velocity are computed by the formula $\delta v^r_j =c(\delta\varphi^1_j-\delta \varphi^2_j)/rB\Delta\theta$, where $\delta\varphi^1_j$ and $\delta\varphi^2_j$ are the floating-potential fluctuations measured by the first and second probes, $r$ is the mean radius of the magnetic flux surface, $c$ is the speed of light, and $B$ is the magnetic field.

Hence, the problem reduces to finding estimates $\hat\alpha, \hat\beta,\hat\lambda$ of parameters of the FSD from measured time sequences $\alpha,\beta,\lambda$ by using the algorithm (\ref{eq:maxLHest}). However, it is incorrect to apply this algorithm to time sequences $\tilde\Gamma_j$. First, we deal with real physical variables that vary in time at a random but finite rate. The finite rate implies that two successive variables in the measure time sample are not statistically independent. Second, an analog-to-digital converter (ADC) introduces additive noise into the measurement results. This noise disturbs homogeneity of the sample and, consequently, we cannot use the algorithm for estimation of the FSD parameters. We have to sample statistically independent variables and to eliminate the noise from the time sequences.

We assume that the observed plasma turbulence is due to the presence of structures that have an enhanced density of charged particles in comparison with the surrounding plasma. We assume that the density of charged particles in a given structure does not depend on the density in the neighboring structures. As a result at the instant at which the structure passes near the probe, its signal is at a maximum, and it is at a minimum when the structure leaves this region. Choosing only maximum and minimum values of the probe current and passing to their increments, we obtain a sample of independent, identically distributed random variables $\Delta\tilde\Gamma'_k=\tilde\Gamma'_k-\tilde\Gamma'_{k+1}, k=1,\dots,N'-1$.

It has been found that the ADC noise does not exceed a certain maximal $C_{\max}$ and minimal $C_{\min}$ levels, which can be determined from a signal measured in the absence of a plasma (before the ECR heating pulse is switched on). The procedure is as follows: before the heating pulse, we determine values $C_{\max}$ and $C_{\min}$ and compute $\Delta C=|C_{\max}-C_{\min}|$. From the sample $\Delta\tilde\Gamma'_k$ we exclude increments that satisfy the condition $|\Delta\tilde\Gamma'|<\Delta C$, and proceed to the new sample $\Delta\tilde\Gamma''_m, m=1,\dots,N''$:
\begin{equation}\label{eq:transfom}
\Delta\tilde\Gamma^*_l=\Delta\tilde\Gamma_{3l-2}''-1/2\left( \Delta\tilde\Gamma_{3l-1}''+\Delta\tilde\Gamma_{3l}''\right),
\end{equation}
where $l=1,\dots,[N''/3]$. Here, $[A]$ is the integer part of $A$ (for more details see \cite{Saenko2009}).

From the sample $\Delta\tilde\Gamma^*_l$ and formulas \eref{eq:maxLHest} we can find estimates $\hat\alpha, \hat\beta, \hat\lambda$ of the parameters $\alpha, \beta, \lambda$.

\section{Results and Conclusions}\label{sec:results}

\begin{figure}[t]
\includegraphics[width=0.47\textwidth]{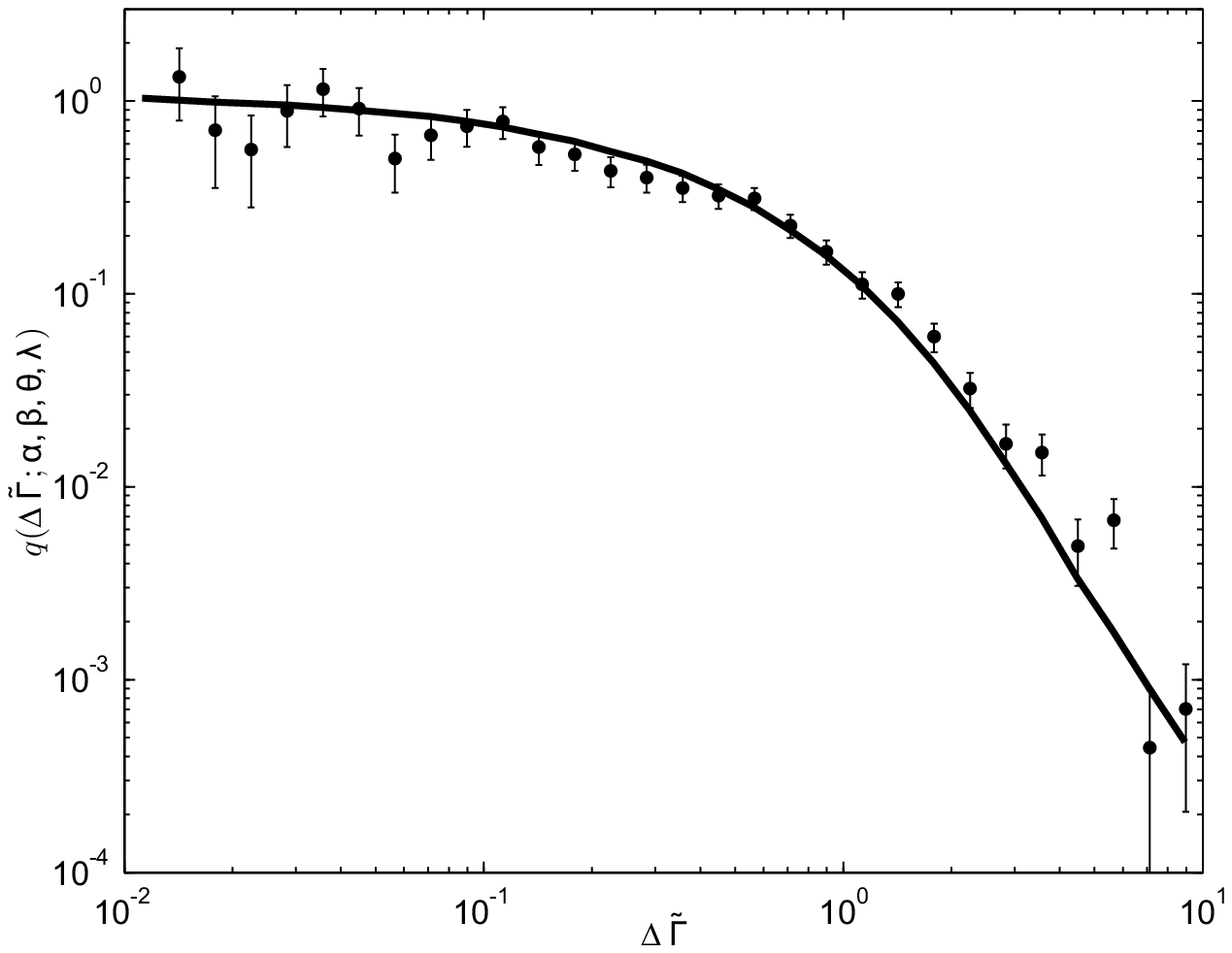}\hfill
\includegraphics[width=0.47\textwidth]{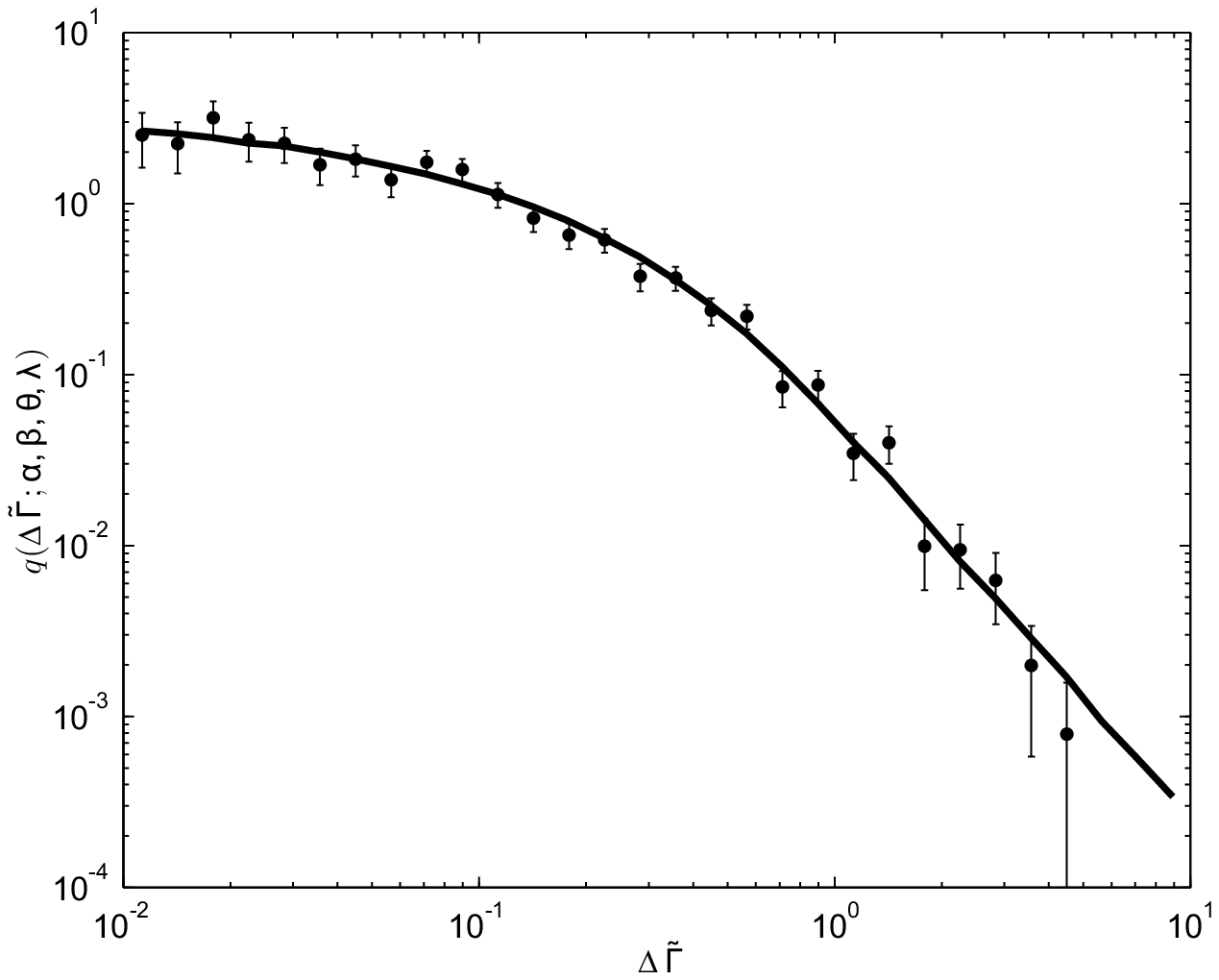}\hfill\\
{\caption{The PDFs of increments of fluctuational fluxes are shown. On the top figure the discharge No. 50659 is
shown and on the bottom figure discharge No. 53080 is shown. \fullcircle are empirical PDF and \full are
FSDs with estimated values of the parameters (see \Tref{tab:estResults}).}\label{fig:flux}}
\end{figure}

The method is described above for approximation of local fluctuational fluxes was used. The fluxes were obtained in the set of experiments  \cite{Meshcheryakov2005} on the device L-2M. The three parameters $\alpha,\beta$ and $\lambda$ were estimated. The results of approximation are presented in the \Tref{tab:estResults}. Also on the \Fref{fig:flux} show the empirical probability densities of increments of amplitudes of turbulent particle fluxes $\Delta\tilde\Gamma^*$. It should be noted that these figures show the increments of amplitudes computed by formula \eref{eq:transfom}, instead of the increments of amplitudes of physical variables under study. It can be seen from these curves that the empirical probability density distributions have tails decreasing as $x^{-\alpha-1}$, which is consistent with the results of other authors \cite{Batanov2002, Jha1992, Carreras1998b, Hidalgo2002a, Carreras1999b, Trasarti-Battistoni2002}. Comparing these distributions with FSDs shows that, within the statistical errors, the FSDs give good fit to the empirical distributions.

\Table{\label{tab:estResults} The estimation results for fluctuation fluxes which were carried out in the work
\cite{Meshcheryakov2005} are presented in the table.}
\br
No. &$\hat\alpha$&$\hat\beta$&$\hat\lambda$&$\chi^2_{e}$&No. &$\hat\alpha$&$\hat\beta$&$\hat\lambda$&$\chi^2_{e}$\\
\mr
50658 &1.54& 0.29&0.70&16.82&53080 &1.28& 0.42&0.20&9.31\\
50659 &1.58& 0.20&0.70&22.07&53125 &1.34& 0.30&0.2&17.47\\
50661 &1.26& 0.50&0.40&10.61&53129 &1.18& 0.62&0.12&11.99\\
50664 &1.31& 0.19&0.50&15.00&53134 &1.38& 0.03&0.20&9.80\\
\br
\endTable

The $\chi^2$-criterion was used to test the hypothesis about coincidence of theoretical and  empirical PDFs. In all cases the following value of the significance level $p=0.01$ and the number of freedom degrees $m=10$ were used. Theoretical value  is $\chi_t^2=23.21$ in all cases. Calculated values of $\chi_e^2$ magnitude in each cases are presented in the \Tref{tab:estResults}. It can be seen that in all cases $\chi_e^2<\chi_t^2$. This  means  that the results don't contradict the hypothesis of the possibility of approximation of local fluctuation fluxes by FSDs.

The FSD, it will be remembered, is a solution to the generalized diffusion equation \eref{eq:commdfs}, hence the coincidence of FSD with the empirical density means that the increments of fluctuation amplitudes obey the generalized diffusion equation. This allows the following conclusions. As is well known, equation \eref{eq:commdfs} describes the asymptotic distribution of the particle coordinate in the CTRW model. This means that stochastic process underlying the turbulent particle fluxes can formally be described in the CTRW model. It should be borne in mind that the walk occurs in the phase space (the time is a characteristic of interest). We use the space  $(t, \tilde\Gamma)$ for \Fref{fig:flux}. Accordingly, the coordinate $x$ in equation \eref{eq:commdfs}  will take  physical variables $\tilde\Gamma$.

\ack
The author is grateful to Yu.V. Khol'nov for providing experimental data and also to G.M. Batanov, K.A. Sarksyan and N.N. Skvortsova for fruitful discussions.
The reported study was partially supported by RFBR, research project No. 10-01-00608 and 12-01-00660, and the Ministry of Education and Science of the Russian Federation (grant 2.1894.2001).

\appendix
\section{Estimation of the parameters by moment method} \label{sec:prmestmom}

Let $Z_1,Z_2,\dots,Z_n$, $n\leqslant4$ be independent, identically distributed random variables with density (\ref{eq:fsd}). The problem is to determine estimates $\hat\alpha, \hat\beta, \hat\theta, \hat\lambda$ of unknown parameters $\alpha,\beta,\theta,\lambda$. This problem was solved in \cite{Bening2004}, where a factional stable stochastic variable was represented in the form
\begin{equation}\label{eq:fsd_rv}
Z(\alpha,\beta,\theta,\lambda)=\lambda\frac{Y(\alpha,\theta)}{[S(\beta,1)]^{\beta/\alpha}}
\end{equation}
where $Y(\alpha,\theta)$ and $S(\beta,1)$ are strictly stable ($0<\alpha\leqslant2$) and one-sided strictly stable ($0<\beta\leqslant1$) stochastic variables with characteristic function (\ref{eq:form_c}).

Here, we only present the final result. The formulas for estimates $\hat\alpha$, $\hat\beta$, $\hat\theta$, $\hat\lambda$ of parameters $\alpha,\beta,\theta,\lambda$ has the form
\begin{eqnarray}
\hat\theta&=1-\frac{2}{n}\sum_{j=1}^n\mathbf{I}(X_j<0),\label{eq:theta_est}\\
\hat\alpha&=\frac{2\pi}{\sqrt{12V_n+\pi^2\left(2Z_n+3\hat\theta^2-1\right)}},\\
\hat\beta&=Z_n\hat\alpha,\\
\hat\lambda&=\exp\left\{U_n-\bbbc(Z_n-1)\right\},\label{eq:lambda_est}
\end{eqnarray}
where $Z_n=\left(1+\frac{M_n}{2\zeta(3)}\right)^{1/3},$ $U_n$, $V_n$, $M_n$ are sample centered logarithmic moments
\begin{eqnarray*}
U_n&=\frac{1}{n}\sum_{j=1}^n\ln|X_j|,
V_n=\frac{1}{n}\sum_{j=1}^n\left(\ln|X_j|-U_n\right)^2,\\
M_n&=\frac{1}{n}\sum_{j=1}^n\left(\ln|X_j|-U_n\right)^3,
\end{eqnarray*}
$\mathbf{I}(A)$ is the indicator of event $A$, $\bbbc=0.5772156649015325\ldots$ is the Eulerian constant, and  $\zeta(3)$ is the Riemann function at point 3.


\end{document}